\documentclass[a4paper]{amsart}

\RequirePackage{amsmath}
\RequirePackage{bm}
\RequirePackage{amssymb}
\RequirePackage{upref}
\RequirePackage{amsthm}
\RequirePackage{enumerate}
\RequirePackage{pb-diagram}
\RequirePackage{amsfonts}
\RequirePackage[mathscr]{eucal}
\RequirePackage{verbatim}
\RequirePackage{xr}
\RequirePackage{graphicx}
\usepackage{calc}
\usepackage{xspace}
\RequirePackage{color}
\RequirePackage{ifthen}

\newcommand{\cf}{cf.\@\xspace}
\newcommand{\resp}{resp.\@\xspace}

\newcommand{\al}{\alpha}

\newcommand{\de}{\delta }

\newcommand{\f}{\varphi}

\newcommand{\ka}{\kappa}
\newcommand{\lam}{\lambda}

\newcommand{\om}{\omega}

\newcommand{\vt}{\vartheta}

\newcommand{\s}{\sigma}

\newcommand{\D}{\varDelta}

\newcommand{\Lam}{\varLambda}
\newcommand{\Om}{\varOmega}

\newcommand{\so}{{\mc S_0}}

\newcommand{\const}{\tup{const}}

\newcommand{\msp[1]}[1]{\mspace{#1mu}}

\newcommand{\R}[1][n+1]{{\protect\mathbb R}^{#1}}

\newcommand{\Ss}[1][n+1]{{\protect\mathbb S}^{#1}}
\newcommand{\Cc}{{\protect\mathbb C}}

\newcommand{\N}{{\protect\mathbb N}}

\newcommand{\Z}{{\protect\mathbb Z}}
\newcommand{\eR}{\stackrel{\lower1ex \hbox{\rule{6.5pt}{0.5pt}}}{\msp[3]\R[]}}
\newcommand{\eN}{\stackrel{\lower1ex \hbox{\rule{6.5pt}{0.5pt}}}{\msp[1]\N}}
\newcommand{\eO}{\stackrel{\lower1ex \hbox{\rule{6pt}{0.5pt}}}{\msc O}}

\newcommand\ra{\rightarrow}

\newcommand{\un}{\infty}
\newcommand{\A}{\forall}

\newcommand{\uu}{\cup}
\newcommand{\ii}{\cap}
\newcommand{\uuu}{\bigcup}

\newcommand{\uud}{ \stackrel{\lower 1ex \hbox {.}}{\uu}}
\newcommand{\uuud}[1]{ \stackrel{\lower 1ex \hbox {.}}{\uuu_{#1}}}
\newcommand\su{\subset}

\newcommand{\sminus}[1][28]{\raise 0.#1ex\hbox{$\scriptstyle\setminus$}}

\newcommand{\abs}[1]{\lvert#1\rvert}

\newcommand{\tit}{\textit}

\newcommand{\tup}{\textup}

\newcommand{\mc}{\protect\mathcal}
\newcommand{\msc}{\protect\mathscr}

\providecommand{\bysame}{\makebox[3em]{\hrulefill}\thinspace}

\newcommand{\bt}{\begin{thm}}
\newcommand{\bl}{\begin{lem}}
\newcommand{\bc}{\begin{cor}}
\newcommand{\bd}{\begin{definition}}
\newcommand{\bpp}{\begin{prop}}
\newcommand{\br}{\begin{rem}}
\newcommand{\bn}{\begin{note}}
\newcommand{\be}{\begin{ex}}
\newcommand{\bes}{\begin{exs}}
\newcommand{\bb}{\begin{example}}
\newcommand{\bbs}{\begin{examples}}
\newcommand{\ba}{\begin{axiom}}
\newcommand{\bas}{\begin{assumption}}

\newcommand{\et}{\end{thm}}
\newcommand{\el}{\end{lem}}
\newcommand{\ec}{\end{cor}}
\newcommand{\ed}{\end{definition}}
\newcommand{\epp}{\end{prop}}
\newcommand{\er}{\end{rem}}
\newcommand{\en}{\end{note}}
\newcommand{\ee}{\end{ex}}
\newcommand{\ees}{\end{exs}}
\newcommand{\eb}{\end{example}}
\newcommand{\ebs}{\end{examples}}
\newcommand{\ea}{\end{axiom}}
\newcommand{\eas}{\end{assumption}}

\newcommand{\bp}{\begin{proof}}
\newcommand{\ep}{\end{proof}}
\newcommand{\eps}{\renewcommand{\qed}{}\end{proof}}

\newcommand{\bal}{\begin{align}}

\newcommand{\bi}[1][1.]{\begin{enumerate}[\upshape #1]}
\newcommand{\bia}[1][(1)]{\begin{enumerate}[\upshape #1]}
\newcommand{\bin}[1][1]{\begin{enumerate}[\upshape\bfseries #1]}
\newcommand{\bir}[1][(i)]{\begin{enumerate}[\upshape #1]}
\newcommand{\bic}[1][(i)]{\begin{enumerate}[\upshape\hspace{2\cma}#1]}
\newcommand{\bis}[2][1.]{\begin{enumerate}[\upshape\hspace{#2\parindent}#1]}
\newcommand{\ei}{\end{enumerate}}

\newcommand\ndots{\raise 0.47ex \hbox {,}\hskip0.06em\cdots %
     \raise 0.47ex \hbox {,}\hskip0.06em} 

\newcommand{\q}{\quad}
\newcommand{\qq}{\qquad}

\newcommand\nd{\noindent}

\newskip\Csmallskipamount                                                
\Csmallskipamount=\smallskipamount
\newskip\Cmedskipamount
\Cmedskipamount=\medskipamount
\newskip\Cbigskipamount
\Cbigskipamount=\bigskipamount

\newcommand\cvs{\vspace\Csmallskipamount}   
\newcommand\cvm{\vspace\Cmedskipamount}

\newskip\csa
\csa=\smallskipamount

\newskip\cma
\cma=\medskipamount

\newskip\cba
\cba=\bigskipamount

\newdimen\spt
\newcommand\citem{\cvs\advance\itemno by
1{(\romannumeral\the\itemno})\hskip3pt}
\newcommand{\bitem}{\cvm\nd\advance\itemno by
1{\bf\the\itemno}\hspace{\cma}}

\newcount\itemno
\newcommand{\lae}[1]{\label{E:#1}}
\newcommand{\lat}[1]{\label{T:#1}}

\newcommand{\re}[1]{\eqref{E:#1}}

\newcommand{\frt}[1]{Theorem~\ref{T:#1} on page~\tup{\pageref{T:#1}}}

\newcommand{\fre}[1]{\eqref{E:#1} on page~\tup{\pageref{E:#1}}}

\newskip\thmskip
\thmskip=\parindent

\newskip\hsk
\newenvironment{hinw}{\labelsep=0pt\begin{list}{}{\labelsep=0pt\itemindent=0pt\labelwidth=0pt\leftmargin=\parindent\rightmargin=0pt\partopsep=\cba}%
\item\it\nopagebreak\nopagebreak}%
{\end{list}}

\newcommand\bh{\begin{hinw}}
\newcommand{\eh}{\end{hinw}}

\newtheoremstyle{normal}
  {\cba}
  {\cba}
  {}
  {\thmskip}
  {\bfseries}
  {.}
  {\hsk}
  {}

\newtheoremstyle{abschnitt}
  {\cba}
  {\cba}
  {}
  {\thmskip}
  {\bfseries}
  {.}
  {\hsk}
  {}

\newtheoremstyle{italic}
  {\cba}
  {\cba}
  {\itshape}
  {\thmskip}
  {\bfseries}
  {.}
  {\hsk}
  {}

\newtheoremstyle{aufgaben}
  {\cba}
  {\cba}
  {}
  {}
  {\normalsize\bfseries}
  {.}
  {\hsk}
  {}

\newtheoremstyle{break}
  {\cba}
  {\cba}
  {\itshape}
  {}
  {\bfseries}
  {.}
  {\newline}
  {}

\swapnumbers
\theoremstyle{italic}
\newtheorem{thm}[subsection]{Theorem}
\newtheorem{lem}[subsection]{Lemma}
\newtheorem{prop}[subsection]{Proposition}
\newtheorem{cor}[subsection]{Corollary}

\theoremstyle{normal}
\newtheorem{rem}[subsection]{Remark}
\newtheorem{definition}[subsection]{Definition}
\newtheorem{example}[subsection]{Example}
\newtheorem{examples}[subsection]{Examples}
\newtheorem{ex}[subsection]{Exercise}
\newtheorem{note}[subsection]{}
\newtheorem{axiom}[subsection]{Axiom}
\newtheorem{assumption}[subsection]{Assumption}

\theoremstyle{aufgaben}
\newtheorem{exs}[subsection]{Exercises}

\swapnumbers

\numberwithin{equation}{section}
\numberwithin{figure}{section}

\newenvironment{textequation}[1][0.8]
{\begin{equation}
\begin{aligned}
\begin{minipage}{#1\linewidth}}
{\end{minipage}
\end{aligned}
\end{equation}
\ignorespacesafterend}

\newcommand{\btext}{\begin{textequation}}
\newcommand{\etext}{\end{textequation}}

\def\hinweis{\@startsection{subsection}{2}%
 \z@{0.7\linespacing\@plus 0.5\linespacing}{0.7\linespacing}%
{\normalfont\itshape\indent}}

\newcounter{hours}\newcounter{minutes}
\newcommand{\printtime}{%
\setcounter{hours}{\time/60}%
\setcounter{minutes}{\time-\value{hours}*60}%
\ifthenelse{\value{minutes}<10}{\thehours :0\theminutes}{\thehours:\theminutes}}

\usepackage[german,english]{babel}
\usepackage{graphicx}
\RequirePackage{amsmath}
\RequirePackage{bm}
\RequirePackage{amssymb}
\RequirePackage{upref}
\RequirePackage{amsthm}
\RequirePackage{enumerate}
\RequirePackage{pb-diagram}
\RequirePackage{amsfonts}
\RequirePackage[mathscr]{eucal}
\RequirePackage{verbatim}
\RequirePackage{xr}
\RequirePackage{graphicx}
\usepackage{calc}
\usepackage{xspace}
\usepackage{dsfont}

\makeatletter
\RequirePackage{color}
\newcommand{\ann}[1]{\renewcommand{\@makefnmark}{\mbox{$^{\color{red}{\@thefnmark}}$}}%
\footnote {#1}}
\makeatother

\RequirePackage{upref}
\RequirePackage{amsthm}
\RequirePackage{enumerate}
\usepackage[mathscr]{eucal}

\usepackage{xr-hyper}

\usepackage{calc}

\newlength{\oddsidemarginlength}
\newlength{\topmarginlength}

\newcounter{numberoflines}
\newcounter{tempcc}
\oddsidemargin=\oddsidemarginlength
\evensidemargin=\oddsidemargin
\topmargin=\topmarginlength
\usepackage[colorlinks=true,linkcolor=blue,citecolor=blue,urlcolor=blue]{hyperref}  

\begin{document}

\flushbottom


\title{The quantization of a black hole}

\author{Claus Gerhardt}
\address{Ruprecht-Karls-Universit\"at, Institut f\"ur Angewandte Mathematik,
Im Neuenheimer Feld 205, 69120 Heidelberg, Germany}
\email{\href{mailto:gerhardt@math.uni-heidelberg.de}{gerhardt@math.uni-heidelberg.de}}
\urladdr{\href{http://www.math.uni-heidelberg.de/studinfo/gerhardt/}{http://www.math.uni-heidelberg.de/studinfo/gerhardt/}}

%
\date{\today}
%


\begin{abstract} 
We apply our model of quantum gravity to an AdS black hole resulting in a wave equation in a quantum spacetime  which has a sequence of solutions that can be expressed as a product of stationary and temporal eigenfunctions. The stationary eigenfunctions can be interpreted as radiation and the temporal as gravitational waves. The event horizon corresponds in the quantum model to a Cauchy hypersurface that can be crossed by causal curves in both directions such that the information paradox does not occur.
\end{abstract}

\maketitle

\tableofcontents

\setcounter{section}{0}
\section{Introduction}
In general relativity the Cauchy development of a Cauchy hypersurface $\so$ is governed by the Einstein equations, where of course the second fundamental form of $\so$ has also to be specified.

 In the model of quantum gravity  we developed in a series of papers \cite{cg:qgravity,cg:uqtheory,cg:uqtheory2,cg:qgravity2, cg:uf2, cg:uf3} we pick a Cauchy hypersurface, which is then only considered to be a complete Riemannian manifold $(\so,g_{ij})$ of dimension $n\ge 3$, and define its quantum development to be the solutions of the wave equation
\begin{equation}\lae{1.1}
\frac1{32}\frac{n^2}{n-1}\Ddot u
-(n-1) t^{2-\frac4n}\D u-\frac {n}2 t^{2-\frac4n}Ru+nt^2\Lam u=0
\end{equation}
defined in the spacetime
\begin{equation}
Q=\so\times (0,\un).
\end{equation}
The Laplacian is the Laplacian with respect to $g_{ij}$, $R$ is the scalar curvature of the metric, $0<t$ is the time coordinate defined by the derivation process of the equation and $\Lam<0$ a cosmological constant. If other physical fields had been present in the Einstein equations then the  equation would contain further lower order terms, \cf \cite{cg:uf2}, but a negative cosmological constant would always have to be present even if the Einstein equations would only describe the vacuum.

Using separation of variables we proved that there is a complete sequence of eigenfunctions $v_j$ (or a complete set of eigendistributions) of a stationary eigenvalue problem and a complete sequence of eigenfunctions $w_i$ of a temporal  implicit eigenvalue problem, where $\Lam$ plays the role of an eigenvalue, such that the functions
\begin{equation}
u(t,x)=w_i(t)v_j(x)
\end{equation}
are solutions of the wave equation, \cf \cite[Section 6]{cg:qgravity2} and \cite{cg:uf2,cg:uf3}.

In this paper we apply this model to quantize an AdS spacetime by picking especially a Cauchy hypersurface in the black hole region of the form
\begin{equation}
\{r=\const<r_0\},
\end{equation}
where $r_0$ is the radius of the event horizon. It turns out that the induced metric of the Cauchy hypersurface can be expressed in the form
\begin{equation}\lae{1.5}
ds^2=d\tau^2+r^2\s_{ij}dx^idx^j,
\end{equation}
where
\begin{equation}
-\un<\tau<\un,
\end{equation}
$r=\const$ and $\s_{ij}$ is the metric of a spaceform $M=M^{n-1}$ with curvature $\tilde \ka$,
\begin{equation}
\tilde\ka\in\{-1,0,1\}.
\end{equation}
The metric in \re{1.5} is free of any coordinate singularity, hence we can let $r$ tend to $r_0$ such that $\so$ represents the event horizon at least topologically. Furthermore, the Laplacian of the metric in \re{1.5} comprises a harmonic oscillator with respect to $\tau$ which enables us to write the stationary eigenfunctions $v_j$ in the form
\begin{equation}
v_j(\tau,x)=\zeta(\tau)\f_j(x),
\end{equation}
where $\f_j$ is an eigenfunction of the Laplacian of $M$ and $\zeta$ a harmonic oscillator the frequency of which are still to be determined.

Due to the presence of the harmonic oscillator we can now consider an \tit{explicit} temporal eigenvalue problem, i.e., we consider the eigenvalue problem
\begin{equation}\lae{1.9}
-\frac1{32} \frac{n^2}{n-1}\Ddot w+n\abs\Lam t^2w=\lam t^{2-\frac4n}w
\end{equation}
with a fixed $\Lam<0$, where we choose $\Lam$ to be the cosmological constant of the AdS spacetime.

The eigenvalue problem \re{1.9} has a complete sequence $(w_i,\lam_i)$ of eigenfunctions with finite energies $\lam_i$ such that
\begin{equation}
0<\lam_0<\lam_1<\cdots
\end{equation}
and by choosing the frequencies of $\zeta$ appropriately we can arrange that the stationary eigenvalues $\mu_j$ of $v_j$ agree with the temporal eigenvalues $\lam_i$. If this is the case then the eigenfunctions
\begin{equation}
u=w_iv_j
\end{equation}
will be a solution of the wave equation. More precisely we proved:
\bt
Let $(\f_j,\tilde\mu_j)$ \resp $(w_i,\lam_i)$ be eigenfunctions of
\begin{equation}
-\tilde\D=-\D_M
\end{equation}
 \resp the temporal eigenfunctions and set
 \begin{equation}
\hat\mu_j=(n-1)r_0^{-2}\tilde\mu_j-\frac n2(n-1)(n-2)r_0^{-2}\tilde\ka.
\end{equation}
Let $\lam_{i_0}$ be the smallest eigenvalue of the $(\lam_i)$ with the property
\begin{equation}
\lam_{i_0}\ge \hat\mu_j,
\end{equation}
then, for any $i\ge i_0$, there exists 
\begin{equation}
\om=\om_{ij}\ge 0
\end{equation}
and corresponding $\zeta_{ij}$ satisfying
\begin{equation}
-\Ddot\zeta_{ij}=r_0^{-2}\om_{ij}^2\zeta_{ij}
\end{equation}
such that
\begin{equation}
\lam_i=\mu_{ij}=(n-1)r_0^{-2}\om_{ij}^2+\hat\mu_j\q\A\, i\ge i_0.
\end{equation}
The functions
\begin{equation}
u_{ij}=w_i\zeta_{ij}\f_j
\end{equation}
are then solutions of the wave equation with bounded energies satisfying
\begin{equation}
\lim_{t\ra 0}w_{ij}(t)=\lim_{t\ra\un}w_{ij}(t)=0
\end{equation}
and
\begin{equation}
w_{ij}\in C^\un(\R[*]_+\times\so)\ii C^{2,\al}(\bar{\R[]}^*_+\times \so)
\end{equation}
for some
\begin{equation}
\frac23\le\al<1.
\end{equation}

Moreover, we have
\begin{equation}
\om_{ij}>0\qq\A\, i>i_0.
\end{equation}
If
\begin{equation}
\lam_{i_0}=\hat\mu_j,
\end{equation}
then we define
\begin{equation}
\zeta_{i_0j}\equiv 1.
\end{equation}
In case $j=0$ and $\tilde\ka\not= -1$  we always have
\begin{equation}
\hat\mu_0\le0
\end{equation}
and 
\begin{equation}
\f_0=\const\not=0
\end{equation}
and hence
\begin{equation}
\om_{i0}>0\qq\A\, i\ge 0.
\end{equation}
\et
\br
(i) The event horizon corresponds to the Cauchy hypersurface $\{t=1\}$ in $Q$ and the open black hole region to the region
\begin{equation}
\so\times (0,1),
\end{equation}
while the open exterior of the black hole region is represented by
\begin{equation}
\so\times (1,\un).
\end{equation}
The black hole singularity corresponds to $\{t=0\}$ which is also a curvature singularity in the quantum spacetime provided we equip $Q$ with a metric such that the hyperbolic operator is normally hyperbolic, \cf \cite[Lemma 6.2]{cg:uf2}. Moreover, in the quantum spacetime the Cauchy hypersurface $\so$ can be crossed by causal curves in both directions, i.e., the information paradox does not occur.

\cvm
(ii) The stationary eigenfunctions can be looked at as being radiation because they comprise the harmonic oscillator, while we consider the temporal eigenfunctions to be gravitational waves.
\er
As it is well-known the Schwarzschild black hole or more specifically the extended Schwarzschild space has already been analyzed by Hawking \cite{hawking:bh} and Hartle and Hawking \cite{hartle-hawking}, see also the book by Wald \cite{wald:qft}, using quantum field theory, but not quantum gravity, to prove that the black hole emits radiation.

\section{The quantization}
The metric in the interior of the black hole can be expressed in the form
\begin{equation}\lae{0.7}
d\bar s^2=-\tilde h^{-1} dr^2 +\tilde hdt^2 + r^2\s_{ij}dx^idx^j,
\end{equation}
where $(\s_{ij})$ is the metric of an $(n-1)$-dimensional space form $M$ and $\tilde h(r)$ is defined by
\begin{equation}
\tilde h=mr^{-(n-2)}+\tfrac2{n(n-1)}\Lam r^2 -\tilde\ka,
\end{equation}
where $m>0$ is the mass of the black hole (or a constant multiple of it),  $\Lam<0$ a cosmological constant, and $\tilde\ka\in\{-1,0,1\}$ is the
curvature of $M=M^{n-1}$, $n\ge 3$. We also stipulate that $M$ is compact in the cases $\tilde\ka\not=1$. If $\tilde\ka =1$ we shall assume
\begin{equation}
M=\Ss[n-1]
\end{equation}
which is of course the important case. By assuming $M$ to be compact we can use eigenfunctions instead of eigendistributions when we consider spatial eigenvalue problems.

The radial variable $r$ ranges between
\begin{equation}
0<r\le r_0,
\end{equation}
where $r_0$ is the radius of the \tit{unique} event horizon.

The interior region of the black hole is a globally hyperbolic $(n+1)$-dimen\-sional spacetime and the hypersurfaces
\begin{equation}
S_r=\{r=\const<r_0\}
\end{equation}
are Cauchy hypersurfaces with induced metric
\begin{equation}
ds^2=\tilde hdt^2 +r^2\s_{ij}dx^idx^j,
\end{equation}
where
\begin{equation}
-\un<t<\un.
\end{equation}
Note that $r=\const$ and hence 
\begin{equation}
0<\tilde h=\const.
\end{equation}
The coordinate transformation
\begin{equation}
\tau=\tilde h^{\frac12}t
\end{equation}
yields
\begin{equation}\lae{2.10}
ds^2=d\tau^2+r^2\s_{ij}dx^idx^j,
\end{equation}
where $\tau\in \R[]$. Since we have removed the coordinate singularity we can now let $r$ converge to $r_0$ such the resulting manifold $\so$ represents the event horizon topologically but with different metric. However, by a slight abuse of language we shall call $\so$ to be Cauchy hypersurface though it is only the geometric limit of Cauchy hypersurfaces.

However, $\so$ is a genuine Cauchy hypersurface in the quantum model which is defined by the equation \fre{1.1}.

Let us now look at the stationary eigenvalue equation
\begin{equation}\lae{2.11}
-(n-1)\D v-\frac n2Rv=\mu v
\end{equation}
in $\so$, where
\begin{equation}
-(n-1)\D v=-(n-1)\Ddot v -(n-1) r_0^{-2}\tilde \D v.
\end{equation}
Using separation of variables let us write
\begin{equation}
v(\tau,x)=\zeta(\tau)\f(x)
\end{equation}
to conclude that the left-hand side of \re{2.11} can be expressed in the form
\begin{equation}
-(n-1)\Ddot\zeta\f+\zeta\{-(n-1)r_0^{-2}\tilde\D\f-\frac n2(n-1)(n-2)r_0^{-2}\tilde\ka\f\},
\end{equation}
since the scalar curvature $R$ of the metric \re{2.10} is
\begin{equation}
R=(n-1)(n-2)r_0^{-2}\tilde\ka.
\end{equation}
Hence, the eigenvalue problem \re{2.11} can be solved by setting
\begin{equation}
v=\zeta\f_j,
\end{equation}
where $\f_j$, $j\in\N$, is an eigenfunction of $-\tilde\D$ such that
\begin{equation}
-\tilde\D\f_j=\tilde\mu_j\f_j,
\end{equation}
\begin{equation}
0=\tilde\mu_0<\tilde\mu_1\le \tilde\mu_2\le \cdots
\end{equation}
and $\zeta$ is an eigenfunction of the harmonic oscillator. The eigenvalue of the harmonic oscillator can be arbitrarily positive or zero. We define it at the moment as
\begin{equation}
r_0^{-2}\om^2
\end{equation}
where $\om\ge 0$ will be determined later. For $\om>0$ we shall consider the real eigenfunction
\begin{equation}\lae{2.23}
\zeta=\sin r_0^{-1}\om\tau
\end{equation}
which represents the ground state in the interval 
\begin{equation}
I_0=(0,\frac\pi{r_0^{-1}\om})
\end{equation}
with vanishing boundary values. $\zeta$ is a solution of the variational problem
\begin{equation}
\frac{\int_{I_0}\abs {\dot\vartheta}^2}{\int_{I_0}\abs\vartheta^2}\ra\min\q\A\, 0\not=\vt\in H^{1,2}_0(I_0)
\end{equation}
in the Sobolev space $H^{1,2}_0(I_0)$.

Multiplying $\zeta$ by a constant we may assume
\begin{equation}
\int_{I_0}\abs\zeta^2=1.
\end{equation}
Obviously,
\begin{equation}
\so=\R[]\times M
\end{equation}
and though $\zeta$ is defined in $\R[]$ and is even an eigenfunction it has infinite norm in $L^2(\R[])$. However, when we consider a finite disjoint union of $N$ open intervals $I_j$
\begin{equation}
\Om=\uuu_{j=1}^NI_j,
\end{equation}
where
\begin{equation}
I_j=(k_j\frac\pi{r_0^{-1}\om},(k_j+1)\frac\pi{r_0^{-1}\om}),\q k_j\in\Z,
\end{equation}
then
\begin{equation}
\zeta_N=N^{-\frac12}\zeta
\end{equation}
is a unit eigenfunction in $\Om$ with vanishing boundary values having the same energy as $\zeta$ in $I_0$. Hence, it suffices to consider $\zeta$ only in $I_0$ since this configuration can immediately be generalized to arbitrary large bounded open intervals
\begin{equation}
\Om\su\R[].
\end{equation}
We then can state:
\bl
There exists a complete sequence of unit eigenfunctions $\f_j$ of $-\tilde\D$ with eigenvalues $\tilde\mu_j$ such that the functions
\begin{equation}
v_j=\zeta \f_j,
\end{equation}
where $\zeta$ is a constant multiple of the function in \re{2.23} with unit $L^2(I_0)$ norm, are solutions of the eigenvalue problem \re{2.11} with eigenvalue
\begin{equation}\lae{2.33}
\mu_j=(n-1)r_0^{-2}\om^2+(n-1)r_0^{-2}\tilde\mu_j-\frac n2 (n-1)(n-2)r_0^{-2}\tilde\ka.
\end{equation}
The eigenfunctions $v_j$ form an orthogonal basis for $L^2(I_0\times M,\Cc)$. The eigenvector
\begin{equation}
v_0=\zeta \f_0,\q \f_0=\const,
\end{equation}
is a ground state with spatial energy
\begin{equation}
(n-1)\int_{I_0\times M}\abs{Dv_0}^2=(n-1)\int_{I_0}\abs{\dot\zeta}^2=(n-1)r_0^{-2}\om^2.
\end{equation}
The energy of the stationary Hamiltonian, i.e., the operator on the left-hand side of \re{2.11}, evaluated at an eigenfunction $v_j$ is equal to the eigenvalue $\mu_j$ in \re{2.33}.
\el

To solve the wave equation \fre{1.1} let us first consider the following eigenvalue problem
\begin{equation}\lae{2.36}
-\frac1{32} \frac{n^2}{n-1}\Ddot w+n\abs\Lam t^2w=\lam t^{2-\frac4n}w
\end{equation}
in the Sobolev space
\begin{equation}
H^{1,2}_0(\R[*]_+).
\end{equation}
Here, 
\begin{equation}
\Lam<0
\end{equation}
can in principle be an arbitrary negative parameter but in the case of an AdS black hole it seems reasonable to choose the cosmological constant of the AdS spacetime. However, if the cosmological constant  is equal to zero, i.e., if we consider a pure Schwarzschild spacetime, then we have either to pick an arbitrary negative constant, if we still want to consider an explicit eigenvalue problem, or we have to consider an implicit eigenvalue problem, where $\Lam$ plays the role of an eigenvalue, \cf \cite[Theorem 6.7]{cg:qgravity2} or \cite[equ. (7.9)]{cg:uf2}. Since our stationary Hamiltonian comprises a harmonic oscillator, the frequency of which is still at our disposal, we would choose a fixed negative $\Lam$, e.g.,
\begin{equation}
\Lam =-1
\end{equation}
 if we wanted to quantize a Schwarzschild black hole.

The eigenvalue problem \re{2.36} can be solved by considering the generalized eigenvalue problem for the bilinear forms
\begin{equation}
B(w,\tilde w)=\int_{\R[*]_+}\{\frac 1{32}\frac{n^2}{n-1}\bar w'\tilde w'+n\abs\Lam t^2\bar w\tilde w\}
\end{equation}
and
\begin{equation}
K(w,\tilde w)=\int_{\R[*]_+}t^{2-\frac4n}\bar w\tilde w
\end{equation}
in the Sobolev space $\mc H$ which is the completion of
\begin{equation}
C^\un_c(\R[*]_+,\Cc)
\end{equation}
in the norm defined by the first bilinear form.

We then look at the generalized eigenvalue problem
\begin{equation}\lae{2.43}
B(w,\f)=\lam K(w,\f)\q\A\,\f\in\mc H
\end{equation}
which is equivalent to \re{2.36}.
\bt\lat{2.2}
The eigenvalue problem \re{2.43} has countably many solutions $(w_i,\lam_i)$ such that
\begin{equation}\lae{2.44}
0<\lam_0<\lam_1<\lam_2<\cdots,
\end{equation}
\begin{equation}
\lim\lam_i=\un,
\end{equation}
and
\begin{equation}
K(w_i,w_j)=\de_{ij}.
\end{equation}
The $w_i$ are complete in $\mc H$ as well as in $L^2(\R[*]_+)$.
\et
\bp
The quadratic form $K$ is compact with respect to the quadratic form $B$ as one can easily prove, \cf \cite[Lemma 6.8]{cg:qfriedman}, and hence a proof of the result, except for the strict inequalities in \re{2.44}, can be found in \cite[Theorem 1.6.3, p. 37]{cg:pdeII}. Each eigenvalue has multiplicity one since we have a linear ODE of order two and all solutions satisfy the boundary condition 
\begin{equation}\lae{6.45}
 w_i(0)=0.
\end{equation}
The kernel is two-dimensional and the condition \re{6.45} defines a one-dimen\-sional subspace. Note, that we considered only real valued solutions to apply this argument. 
\ep
We are now ready to define the solutions of the wave equation \re{1.1}.
\bt
Let $(\f_j,\tilde\mu_j)$ \resp $(w_i,\lam_i)$ be eigenfunctions of
\begin{equation}
-\tilde\D=-\D_M
\end{equation}
 \resp the temporal eigenfunctions and set
 \begin{equation}
\hat\mu_j=(n-1)r_0^{-2}\tilde\mu_j-\frac n2(n-1)(n-2)r_0^{-2}\tilde\ka.
\end{equation}
Let $\lam_{i_0}$ be the smallest eigenvalue of the $(\lam_i)$ with the property
\begin{equation}
\lam_{i_0}\ge \hat\mu_j,
\end{equation}
then, for any $i\ge i_0$, there exists 
\begin{equation}
\om=\om_{ij}\ge 0
\end{equation}
and corresponding $\zeta_{ij}$ satisfying
\begin{equation}
-\Ddot\zeta_{ij}=r_0^{-2}\om_{ij}^2\zeta_{ij}
\end{equation}
such that
\begin{equation}
\lam_i=\mu_{ij}=(n-1)r_0^{-2}\om_{ij}^2+\hat\mu_j\q\A\, i\ge i_0.
\end{equation}
The functions
\begin{equation}
u_{ij}=w_i\zeta_{ij}\f_j
\end{equation}
are then solutions of the wave equation with bounded energies satisfying
\begin{equation}
\lim_{t\ra 0}w_{ij}(t)=\lim_{t\ra\un}w_{ij}(t)=0
\end{equation}
and
\begin{equation}
w_{ij}\in C^\un(\R[*]_+\times\so)\ii C^{2,\al}(\bar{\R[]}^*_+\times \so)
\end{equation}
for some
\begin{equation}
\frac23\le\al<1.
\end{equation}

Moreover, we have
\begin{equation}
\om_{ij}>0\qq\A\, i>i_0.
\end{equation}
If
\begin{equation}
\lam_{i_0}=\hat\mu_j,
\end{equation}
then we define
\begin{equation}
\zeta_{i_0j}\equiv 1.
\end{equation}
In case $j=0$ and $\tilde\ka\not= -1$  we always have
\begin{equation}
\hat\mu_0\le 0
\end{equation}
and 
\begin{equation}
\f_0=\const\not=0
\end{equation}
and hence
\begin{equation}
\om_{i0}>0\qq\A\, i\ge 0.
\end{equation}
\et
\bp
The proof is obvious.
\ep
\br
(i) By construction the temporal and spatial energies of the solutions of the wave equation have to be equal.

\cvm
(ii) The stationary solutions comprising a harmonic oscillator can be looked at a being radiation while we consider the temporal solutions to be gravitational waves.

\cvm
(iii) If one wants to replace the bounded Interval $I_0$ by $\R[]$ then the eigenfunctions $\zeta_{ij}$ have to be replaced by eigendistributions. An appropriate choice would be
\begin{equation}
\zeta_{ij}=e^{ir_0^{-1}\om_{ij}\tau};
\end{equation}
also see \cite{cg:uf3} for a more general setting.
\er

The hyperbolic operator defined by the wave equation \fre{1.1} can be defined in the spacetime
\begin{equation}
Q=\so\times\R[*]_+
\end{equation}
which can be equipped with the Lorentzian metrics
\begin{equation}\lae{6.17}
d\bar s^2=-\frac{32(n-1)}{n^2}dt^2+g_{ij}dx^idx^j
\end{equation}
as well as with the metric
\begin{equation}\lae{6.18}
d\tilde s^2=-\frac{32(n-1)}{n^2}dt^2+\frac1{n-1}t^{\frac4n-2}g_{ij}dx^idx^j,
\end{equation}
where $g_{ij}$ is the metric defined on $\so$ and the indices now have the range $1\le i,j\le n$. In both metrics $Q$ is globally hyperbolic provided $\so$ is complete, which is the case for the metric defined in \re{2.10}. The hyperbolic operator is symmetric in the first metric but not normally hyperbolic while it is normally hyperbolic but not symmetric in the second metric. Normally hyperbolic means that the main part of the operator is identical to the Laplacian of the spacetime metric.

Hence, if we want to describe quantum gravity not only by an equation but also by the metric of a spacetime then the metric in \re{6.18} has to be chosen. In this metric $Q$ has a curvature singularity in $t=0$, \cf \cite[Remark 6.3]{cg:uf2}. The Cauchy hypersurface $\so$ then corresponds to the hypersurface
\begin{equation}
\{t=1\}
\end{equation}
which also follows from the derivation of the quantum model where we consider a fiber bundle  $E$ with base space $\so$ and the elements of the fibers were Riemann metrics of the form
\begin{equation}
g_{ij}(t,x)=t^\frac4n\s_{ij}(x)
\end{equation}
where $\s_{ij}$ were  metrics defined in $\so$ and $t$ is the time coordinate that we use in $Q$, i.e.,
\begin{equation}
g_{ij}(1,x)=\s_{ij}(x).
\end{equation}
In the present situation we used the symbol $g_{ij}$ to denote the metric on $\so$ since $\s_{ij}$ is supposed to be the metric of the spaceform $M$.

Thus the event horizon is characterized by the Cauchy hypersurface
\begin{equation}
\{t=1\}
\end{equation}
and obviously we shall assume that the black hole singularity
\begin{equation}
\{r=0\}
\end{equation}
corresponds to the curvature singularity
\begin{equation}
\{t=0\}
\end{equation}
of $Q$, i.e., the open black hole region is described in the quantum model by
\begin{equation}
\so\times (0,1)
\end{equation}
and the open exterior region by
\begin{equation}
\so\times (1,\un).
\end{equation}
Stipulating that the time orientation in the quantum model should be the same as in the AdS spacetime we conclude that the curvature singularity $t=0$ is a future singularity, i.e, the present time function is not future directed. To obtain a future directed coordinate system we have to choose $t$ negative, i.e.,
\begin{equation}
Q=\so\times (-\un,0).
\end{equation}
In the metric \re{6.18} we then have to replace
\begin{equation}
t^{\frac4n-2}
\end{equation}
by
\begin{equation}\lae{2.78}
\abs t^{\frac4n-2}
\end{equation}
and similarly in the wave equation, which is then invariant with respect to the reflection
\begin{equation}
t\ra -t.
\end{equation}
As a final remark in this section let us state:
\br
In the quantum model of the black hole the event horizon is a regular Cauchy hypersurface and  can be crossed in both directions by causal curves hence no information paradox can occur.
\er

\section{Transition from the black hole to the white hole}

We shall choose the time variable $t$ negative to have a future oriented coordinate system. The quantum model of the white hole will then be described by a positive $t$ variable and the transition from black to white hole would be future oriented. Obviously, we only have to consider the temporal eigenvalue equation \fre{2.36} to define a transition, where of course \fre{2.78} and its inverse relation have to be observed.

Since the coefficients of the ODE in \re{2.36} are at least H\"older continuous in $\R[]$, a solution $w$ defined on the negative axis has a natural extension to $\R[]$ since we know that $w(0)$=0. Denote the fully extended function by $w$ too, then
\begin{equation}
w\in C^{2,\al}(\R[]),
\end{equation}
where we now, without loss of generality,  only consider a real solution.
\bt
A naturally extended solution $w$ of  the temporal eigenvalue equation \re{2.36} is antisymmetric in $t$,
\begin{equation}\lae{3.2}
w(-t)=-w(t)
\end{equation}
and the restriction of $w$ to the positive axis is also a variational solution as defined in \frt{2.2}.
\et
\bp
It suffices to prove \re{3.2}. Let $t>0$ and define
\begin{equation}
\tilde w(t)=-w(-t),
\end{equation}
then $\tilde w$ solves the ODE in $\R[*]_+$ and
\begin{equation}
\tilde w(0)=w(0)=0
\end{equation}
as well as
\begin{equation}
\dot{\tilde w}(0)=\dot w(0),
\end{equation}
hence we deduce
\begin{equation}
\tilde w(t)=w(t)\qq\A\, t>0
\end{equation}
because the solutions of a second order ODE are uniquely determined by the initial values of the function and its derivative. 
\ep

\br
This transition result is also valid in the general case when the curvature singularity in $t=0$ does not necessarily correspond to the singularity of a black or white hole. The quantum evolution of any Cauchy hypersurface $\so$ in a globally hyperbolic spacetime will always have a curvature singularity either in the past or in the future of $\so$ and the evolution can be extended past this singularity. Note also that the quantum Lorentzian  distance to that past or future singularity is always finite.
\er

\bibliographystyle{hamsplain}

\begin{thebibliography}{10}

\bibitem{cg:qfriedman}
Claus Gerhardt, \emph{{Quantum cosmological Friedman models with an initial
  singularity}}, Class. Quantum Grav. \textbf{26} (2009), no.~1, 015001,
  {\href{http://arXiv.org/abs/0806.1769}{arXiv:0806.1769}},
  {\href{http://dx.doi.org/10.1088/0264-9381/26/1/015001}{doi:10.1088/0264-9381/26/1/015001}}.

\bibitem{cg:uqtheory2}
\bysame, \emph{{A unified quantum theory II: gravity interacting with
  Yang-Mills and spinor fields}}, 2013,
  {\href{http://arXiv.org/abs/1301.6101}{arXiv:1301.6101}}.

\bibitem{cg:pdeII}
\bysame, \emph{{Partial differential equations II}}, Lecture Notes, University
  of Heidelberg, 2013,
  {\href{http://www.math.uni-heidelberg.de/studinfo/gerhardt/PDE2.pdf}{pdf
  file}}.

\bibitem{cg:qgravity}
\bysame, \emph{{The quantization of gravity in globally hyperbolic
  spacetimes}}, Adv. Theor. Math. Phys. \textbf{17} (2013), no.~6, 1357--1391,
  {\href{http://arXiv.org/abs/1205.1427}{arXiv:1205.1427}},
  {\href{http://dx.doi.org/10.4310/ATMP.2013.v17.n6.a5}{doi:10.4310/ATMP.2013.v17.n6.a5}}.

\bibitem{cg:uqtheory}
\bysame, \emph{{A unified quantum theory I: gravity interacting with a
  Yang-Mills field}}, Adv. Theor. Math. Phys. \textbf{18} (2014), no.~5,
  1043--1062, {\href{http://arXiv.org/abs/1207.0491}{arXiv:1207.0491}},
  {\href{http://dx.doi.org/10.4310/ATMP.2014.v18.n5.a2}{doi:10.4310/ATMP.2014.v18.n5.a2}}.

\bibitem{cg:qgravity2}
\bysame, \emph{{A unified field theory I: The quantization of gravity}},
  (2015), {\href{http://arXiv.org/abs/1501.01205}{arXiv:1501.01205}}.

\bibitem{cg:uf2}
\bysame, \emph{{A unified field theory II: Gravity interacting with a
  Yang-Mills and Higgs field}},  (2016),
  {\href{http://arXiv.org/abs/1602.07191}{arXiv:1602.07191}}.

\bibitem{cg:uf3}
\bysame, \emph{{Deriving a complete set of eigendistributions for a
  gravitational wave equation describing the quantized interaction of gravity
  with a Yang-Mills field in case the Cauchy hypersurface is non-compact}},
  (2016), {\href{http://arXiv.org/abs/1605.03519}{arXiv:1605.03519}}.

\bibitem{hartle-hawking}
J.~B. Hartle and S.~W. Hawking, \emph{Path-integral derivation of black-hole
  radiance}, Phys. Rev. D \textbf{13} (1976), no.~8, 2188--2203,
  {\href{http://dx.doi.org/10.1103/PhysRevD.13.2188}{doi:10.1103/PhysRevD.13.2188}}.

\bibitem{hawking:bh}
S.W. Hawking, \emph{{Particle creation by black holes}}, Communications in
  Mathematical Physics \textbf{43} (1975), no.~3, 199--220,
  {\href{http://dx.doi.org/10.1007/BF02345020}{doi:10.1007/BF02345020}}.

\bibitem{wald:qft}
Robert~M. Wald, \emph{Quantum field theory in curved spacetime and black hole
  thermodynamics}, Chicago Lectures in Physics, University of Chicago Press,
  Chicago, IL, 1994.

\end{thebibliography}
\providecommand{\bysame}{\leavevmode\hbox to3em{\hrulefill}\thinspace}
\providecommand{\href}[2]{#2}



\end{document}